\documentclass[12pt]{article}
\usepackage{amssymb}
\usepackage{amsmath}
\usepackage{color}
\usepackage{euscript}
\def\d{\delta}

\def\la{\lambda}

\def\be{\begin{equation}}
\def\ee{\end{equation}}
\def\arr{\begin{array}{rll}}
\def\ea{\end{array}}
\def\bea{\begin{eqnarray}}
\def\eea{\end{eqnarray}}

\def\N2{$N{=}2$}

\def\>{\rangle}
\def\<{\langle}
\def\+{\dagger}
\def\={\ =\ }

\begin{document}
\renewcommand{\thefootnote}{\arabic{footnote}}
\noindent
\begin{titlepage}
\setcounter{page}{0}
\begin{flushright}
$\qquad$
\end{flushright}
\vskip 3cm
\begin{center}
{\LARGE\bf{$\mathcal{N}=2$ supersymmetric odd-order}
\vskip 0.5cm
{\LARGE\bf Pais-Uhlenbeck oscillator}}
\vskip 1cm
$
\textrm{\Large Ivan Masterov\ }
$
\vskip 0.7cm
{\it
Laboratory of Mathematical Physics, Tomsk Polytechnic University, \\
634050 Tomsk, Lenin Ave. 30, Russian Federation}
\vskip 0.7cm
{E-mail: masterov@tpu.ru}

\end{center}
\vskip 1cm
\begin{abstract}
\noindent
We consider an $\,\mathcal{N}=2$ supersymmetric odd-order Pais-Uhlenbeck oscillator with distinct frequencies of oscillation. The technique previously developed in [Acta Phys. Polon. B 36 (2005) 2115; Nucl. Phys. B 902 (2016) 95] is used to construct a family of Hamiltonian structures for this system.
\end{abstract}


\vskip 0.5cm


\end{titlepage}
\noindent
{\bf{\large 1. Introduction}}
\vskip 0.5cm
\noindent
A systematic way to construct a Hamiltonian formulation for nondegenerate higher-derivative mechanical systems is based on Ostrogradsky's approach \cite{Ostrogradski}. Canonical formalism for degenerate higher-derivative models can be obtained with the aid of Dirac's method for constrained systems \cite{Dirac} or by applying the Faddeev-Jackiw prescription \cite{Faddeev}.

However, some higher-derivative models are multi-Hamiltonian. The simplest example of such systems is the one-dimensional fourth-order Pais-Uhlenbeck (PU) oscillator \cite{Pais}. Ostrogradsky's Hamiltonian of this system is unbounded from below. As a consequence, quantum theory of the model faces ghost-problem (see, e.g., a detailed discussion in Ref. \cite{Woodard}). For distinct frequencies of oscillation, this Hamiltonian can be presented as a difference of two harmonic oscillators by applying an appropriate canonical transformation \cite{Pais,Smilga}.
This representation provides two functionally independent positive-definite integrals of motion.
As was observed in \cite{Kosinski} (see also Ref. \cite{KLS}), a linear combination involving arbitrary nonzero coefficients of these constants of motion can also play a role of a Hamiltonian for the fourth-order PU oscillator\footnote{An alternative Hamiltonian formulation for the fourth-order PU oscillator has been also constructed in paper \cite{Ali_1}.}.
Thus, for positive coefficients, the alternative Hamiltonian is positive-definite and consequently is more relevant for quantization than Ostrogradsky's one.

For arbitrary odd and even orders, the PU oscillator with distinct frequencies of oscillation can also be treated by the technique employed in Ref. \cite{Kosinski}. This fact has been established in \cite{Alt,Alt_1} (see also Ref. \cite{KL}) where the corresponding families of Hamiltonian structures have been constructed.
The main advantage of the alternative Hamiltonian formulation obtained in such a way is that this may correspond to a positive-definite Hamiltonian.

The even-order PU oscillator with distinct frequencies of oscillation admits an $\,\mathcal{N}=2$ supersymmetric extension \cite{Masterov_N2PU}.
This generalization is invariant under the time translations. However, the Noether charge associated with this symmetry can be presented as a sum of $\,\mathcal{N}=2$ supersymmetric harmonic oscillators which alternate in a sign \cite{Masterov_N2PU} (see also Ref. \cite{Smilga_1}). A canonical formalism with regard to a such Hamiltonian brings about trouble with ghosts upon quantization \cite{Masterov_N2PU}. This problem is reflected in the fact that the quantum state space of the model contains negative norm states, while a ground state is absent. In Ref. \cite{Alt} an alternative Hamiltonian formulation for an $\,\mathcal{N}=2$ supersymmetric even-order PU oscillator has been constructed so as to avoid these nasty features.

For a particular choice of oscillation frequencies, an $\,\mathcal{N}=2$ supersymmetric extension of the odd-order PU oscillator has been derived in Ref. \cite{Masterov_1}. It has been shown that this extension accommodates conformal symmetry provided frequencies of oscillation form a certain arithmetic sequence. Any other aspects related with the $\,\mathcal{N}=2$ supersymmetric odd-order PU oscillator remain completely unexplored. In particular, a canonical formulation of this model has not yet been considered. The purpose of the present work is to construct a Hamiltonian formulation for an $\,\mathcal{N}=2$ supersymmetric odd-order PU oscillator with distinct frequencies of oscillation by applying the technique previously developed in Refs. \cite{Kosinski,Alt}.

The paper is organized as follows. In the next section we consider the odd-order PU oscillator with distinct frequencies of oscillation and introduce an $\mathcal{N}=2$ supersymmetric extension of this model. A Hamiltonian formulation for an $\,\mathcal{N}=2$ supersymmetric third-order PU oscillator is constructed in Sect. 3, while the general case is treated in Sect. 4. In Sect. 5, a quantum version of the $\,\mathcal{N}=2$ supersymmetric odd-order PU oscillator is considered. We summarize our results and discuss further possible developments in the concluding Sect. 6. Some technical details are given in Appendix. Throughout the work summation over repeated spatial indices is understood, unless otherwise is explicitly stated. Both a superscript in braces and a number of dots over spatial coordinates designate the number of derivatives with respect to time. Complex conjugation of a function $f$ is denoted by $f^*$. Hermitian conjugation of an operator $\hat{a}$ is designated as $(\hat{a})^\dagger$.

\vskip 0.5cm
\noindent
{\bf{\large 2. The model}}
\vskip 0.5cm
Symmetries of the PU oscillator have recently attracted some attention \cite{Gomis}-\cite{Andr}. The interest was motivated by the desire to realize the so-called $l$-conformal Newton-Hooke algebra \cite{Negro_1}-\cite{Galajinsky_3} in this model. As was shown in \cite{Masterov_2} (see also Ref. \cite{Liu}), the $(2n+1)$-order PU oscillator, which accommodates the $l$-conformal Newton-Hooke symmetry, is described by the action functional\footnote{Some aspects of the third-order PU oscillator have been studied in \cite{Lukier_1}-\cite{Lukier_2} (see also Refs. \cite{Horvathy_1}-\cite{Gomis_1}).}
\bea\label{odd_conf}
S=\frac{1}{2}\int\,dt\,\epsilon_{ij}\,x_i\prod_{k=1}^{n}\left(\frac{d^2}{dt^2}+k^2\omega^2\right)\dot{x}_j,
\eea
where $\epsilon_{ij}$ is the Levi-Civit\'{a} symbol with $\epsilon_{12}=1$.

Symmetry structure intrinsic to the model (\ref{odd_conf}) allows one to construct an $\,\mathcal{N}=2$ supersymmetric generalization of this system with the aid of Niederer-like coordinate transformations \cite{Masterov_1,Niederer}. The action functional associated with this extension reads
\bea\label{1}
\begin{aligned}
&
S=\frac{1}{2}\int dt\,\epsilon_{ij}\left(x_i\prod_{k=1}^{n}\left(\frac{d^2}{dt^2}+k^2\omega^2\right)\dot{x}_j
-\psi_i\left(\frac{d}{dt}+in\omega\right)\prod_{k=1}^{n-1}\left(\frac{d^2}{dt^2}+k^2\omega^2\right)\dot{\bar{\psi}}_j-\right.
\\[2pt]
&
\left.\qquad\qquad\;-\bar{\psi}_i\left(\frac{d}{dt}-in\omega\right)\prod_{k=1}^{n-1}\left(\frac{d^2}{dt^2}+k^2\omega^2\right)\dot{\psi}_j-
z_i\prod_{k=1}^{n-1}\left(\frac{d^2}{dt^2}+k^2\omega^2\right)\dot{z}_j\right).
\end{aligned}
\eea
The configuration space of this model involves the real bosonic coordinates $x_i$, the fermionic coordinates $\psi_i$, $\bar{\psi}_i$, which are complex conjugates of each other $\bar{\psi}_i=\psi_i^*$, and real extra bosonic coordinates $z_i$. The model (\ref{1}) is invariant under the supersymmetry transformations of the form \cite{Masterov_1}
\bea\label{3}
\begin{aligned}
&
\d x_i=i\psi_i\alpha+i\bar{\psi}_i\bar{\alpha},\qquad \d z_i=\left(-\dot{\psi}_i+in\omega\psi_i\right)\alpha+
\left(\dot{\bar{\psi}}_i+in\omega\bar{\psi}_i\right)\bar{\alpha},
\\[2pt]
&
\quad\;\d\psi_i=\left(\dot{x}_i+in\omega x_i-iz_i\right)\bar{\alpha},\qquad\d\bar{\psi}_i=\left(\dot{x}_i-in\omega x_i+i z_i\right)\alpha,
\end{aligned}
\eea
where $\alpha$ and $\bar{\alpha}$ are odd infinitesimal parameters.

It is evident that the model (\ref{odd_conf}) can be generalized to the case of arbitrary distinct oscillation frequencies. For this purpose, the action (\ref{odd_conf}) can be transformed to the form \cite{Alt_1}
\bea\label{model}
S=\frac{1}{2}\int dt\,\epsilon_{ij}\, x_i\prod_{k=0}^{n-1}\left(\frac{d^2}{dt^2}+\omega_k^2\right)\dot{x}_j.
\eea
For definiteness, we assume that $0<\omega_0<\omega_1<..<\omega_{n-1}$. On the other hand, a possibility to generalize the model (\ref{1}) along similar lines is less obvious, because we must simultaneously change both the action functional (\ref{1}) and the supersymmetry transformations (\ref{3}). By analogy with the analysis in Ref. \cite{Masterov_N2PU}, let us abandon the conformal invariance and modify the action functional (\ref{1}) as follows
\bea\label{action}
\begin{aligned}
&
S=\frac{1}{2}\int dt\,\epsilon_{ij}\left(x_i\prod_{k=0}^{n-1}\left(\frac{d^2}{dt^2}+\omega_k^2\right)\dot{x}_j- i\psi_i\prod_{k=-n+1}^{n-1}\left(\frac{d}{dt}+i\omega_k\right)\dot{\bar{\psi}}_j-\right.
\\[2pt]
&
\left.\qquad\qquad\quad\,-i\bar{\psi}_i\prod_{k=-n+1}^{n-1}\left(\frac{d}{dt}-i\omega_k\right)\dot{\psi}_j
-z_i\prod_{k=1}^{n-1}\left(\frac{d^2}{dt^2}+\omega_k^2\right)\dot{z}_j\right),
\end{aligned}
\eea
where, for convenience, we denoted $\omega_{-k}=-\omega_k$.  The dynamics of this model is governed by the equations of motion
\bea
\sum_{k=0}^{n}\sigma_k^{n,0} x_i^{(2k+1)}=0,\, \sum_{k=0}^{2n-1}(-i)^{2n-k-1}\sigma_k^{n}\psi_i^{(k+1)}=0,\, \sum_{k=0}^{2n-1}i^{2n-k-1}\sigma_k^{n}\bar{\psi}_i^{(k+1)}=0,\,
\sum_{k=0}^{n-1}\sigma_k^{n,1} z_i^{(2k+1)}=0,
\nonumber
\eea
where $\sigma_{k}^{n,s}$, $\sigma_{k}^{n}$ are elementary symmetric polynomials defined by\footnote{By definition, we put $\sigma_{n-s}^{n,s}\equiv 1$ for $s=0,1$, $\sigma_{2n-1}^n\equiv 1$.}
\bea
\begin{aligned}
&
\sigma_{k}^{n,s}=\sum_{i_1<i_2<..<i_{n-k}=s}^{n-1}\omega_{i_1}^2\omega_{i_2}^2 ...\omega_{i_{n-k}}^2,\qquad
\sigma_k^{n}=\sum_{i_1<i_2<..<i_{2n-k-1}=-n+1}^{n-1}\omega_{i_1}\omega_{i_2}...\omega_{i_{2n-k-1}}.
\end{aligned}
\nonumber
\eea

Let us show that the model (\ref{action}) is an $\,\mathcal{N}=2$ supersymmetric extension of the odd-order PU oscillator (\ref{model}). As the first step, one finds the Noether charge which corresponds to the invariance of the model (\ref{action}) under the time translations
\bea\label{H}
\begin{aligned}
&
H=\sum_{k=1}^{n}\sum_{m=0}^{n-k}\sigma_{k+m}^{n,0}\epsilon_{ij}x_i^{(2k)}x_j^{(2m+1)}-
\sum_{k=1}^{n-1}\sum_{m=0}^{n-k-1}\sigma_{k+m}^{n,1}\epsilon_{ij}z_i^{(2k)}z_j^{(2m+1)}+
\\[2pt]
&
\qquad\qquad+(-1)^{n+1}\sum_{k=1}^{2n-1}\sum_{m=0}^{2n-k-1}i^{k-m}\sigma^{n}_{k+m}\epsilon_{ij}\psi_i^{(k)}\bar{\psi}_j^{(m+1)}.
\end{aligned}
\eea
Dirac's Hamiltonian of the system (\ref{action}) is the phase space analogue of this conserved quantity. Therefore, there exists such a graded Poisson bracket $[\cdot,\cdot\}$ that the relations
\bea\label{equ}
\begin{aligned}
&
[x_i^{(k)},H\}=x_i^{(k+1)},&& k=0,1,..,2n-1,&& [x_i^{(2n)},H\}=-\sum_{k=0}^{n-1}\sigma_k^{n,0}x_i^{(2k+1)},
\\[2pt]
&
[\psi_i^{(k)},H\}=\psi_i^{(k+1)},&& k=0,1,..,2n-2,&& [\psi_i^{(2n-1)},H\}=-\sum_{k=0}^{2n-2}(-i)^{2n-k-1}\sigma_k^{n}\psi_i^{(k+1)},
\\[2pt]
&
[\bar\psi_i^{(k)},H\}=\bar\psi_i^{(k+1)},&& k=0,1,..,2n-2,&& [\bar\psi_i^{(2n-1)},H\}=-\sum_{k=0}^{2n-2}i^{2n-k-1}\sigma_k^{n}\bar\psi_i^{(k+1)},
\\[2pt]
&
[z_i^{(k)},H\}=z_i^{(k+1)},&& k=0,1,..,2n-3,&& [z_i^{(2n-2)},H\}=-\sum_{k=0}^{n-2}\sigma_k^{n,1}z_i^{(2k+1)},
\end{aligned}
\eea
hold. It is straightforward to verify (for some technical details see Appendix) that this bracket can be defined as follows\footnote{For the model (\ref{model}), an analogue of the bracket (\ref{PB}) has been introduced in Ref. \cite{Alt_1}}
\bea\label{PB}
\begin{aligned}
&
[A,B\}=(-1)^{n+1}\left(\sum_{r,m=0}^{2n}\mu_{rm}^{n,0}\epsilon_{ij}\frac{\partial A}{\partial x_i^{(r)}}\frac{\partial B}{\partial x_j^{(m)}}
+\sum_{r,m=0}^{2n-2}\mu_{rm}^{n,1}\epsilon_{ij}\frac{\partial A}{\partial z_i^{(r)}}\frac{\partial B}{\partial z_j^{(m)}}+\right.
\\[2pt]
&
\left.\qquad\qquad-\sum_{r,m=0}^{2n-1}\upsilon_{rm}\epsilon_{ij}\frac{\overleftarrow{\partial}A}{\partial\psi_i^{(r)}}
\frac{\overrightarrow{\partial B}}{\partial\bar{\psi}_j^{(m)}}
+\sum_{r,m=0}^{2n-1}(\upsilon_{rm})^*\epsilon_{ij}\frac{\overleftarrow{\partial}A}{\partial\bar\psi_i^{(r)}}
\frac{\overrightarrow{\partial B}}{\partial\psi_j^{(m)}}\right),
\end{aligned}
\eea
where the coefficients $\mu_{rm}^{n,0}$, $\mu_{rm}^{n,1}$, and $\upsilon_{rm}$ are given by
\bea\label{Pois_2}
\mu_{rm}^{n,s}=\left\{
\begin{aligned}
&
0
\\[2pt]
&
(-1)^{\frac{r-m}{2}}P_{r+m-2n+2s}^{n,s}
\end{aligned},\;
\upsilon_{rm}=\left\{
\begin{aligned}
&
i^{r-m}P_{r+m-2n+1}^{n,0},&& r+m\,-\,\mbox{odd}
\\[2pt]
&
\omega_0 i^{r-m} P_{r+m-2n}^{n,0}, && r+m\,-\,\mbox{even}
\end{aligned}
\right.
\right.,
\nonumber
\eea
with $P_{2k}^{n,s}$ being the $k$-th degree symmetric polynomial in $(n-s)$ variables $\omega_s^2$, $\omega_{s+1}^2$,.., $\omega_{n-1}^2$
\bea\label{P}
P_{2k}^{n,s}=\sum_{\la_{s},\la_{s+1},..,\la_{n-1}=0 \atop\la_s+\la_{s+1}+..+\la_{n-1}=k}^{k}\omega_s^{2\la_s}\omega_{s+1}^{2\la_{s+1}}...
\omega_{n-1}^{2\la_{n-1}}.
\nonumber
\eea
By definition, this polynomial is equal to zero for negative values of $k$. In the next sections we will show that (\ref{PB}) possesses the standard properties of a graded Poisson bracket.

As the next step, we need to generalize the supersymmetry transformations (\ref{3}) to the case of arbitrary distinct oscillation frequencies. To this end, let us note that the transformations (\ref{3}) are also available for an $\,\mathcal{N}=2$ supersymmetric even-order PU oscillator which exhibits conformal invariance \cite{Masterov_1}. Therefore, it is natural to expect that both the model (\ref{action}) and its even-order analogue \cite{Masterov_N2PU} are invariant with respect to the supersymmetry transformations
\bea\label{2}
\begin{aligned}
&
\d x_i=\psi_i\alpha+\bar{\psi}_i\bar{\alpha},\qquad \d z_i=\left(i\dot{\psi}_i+\omega_0\psi_i\right)\alpha+
\left(-i\dot{\bar{\psi}}_i+\omega_0\bar{\psi}_i\right)\bar{\alpha},
\\[2pt]
&
\;\;\;\d\psi_i=\left(-i\dot{x}_i+\omega_0 x_i-z_i\right)\bar{\alpha},\qquad\d\bar{\psi}_i=\left(-i\dot{x}_i-\omega_0 x_i+ z_i\right)\alpha,
\end{aligned}
\eea
which have been introduced in Ref. \cite{Masterov_N2PU} for an $\,\mathcal{N}=2$ supersymmetric even-order PU oscillator. It is straightforward to verify that this is the case. The integrals of motion, which correspond to these transformations, read
\bea\label{Q}
&&
Q=-\sum\limits_{k=1}^{n-1}\sum\limits_{m=0}^{n-k-1}\sigma_{k+m}^{n,1}\epsilon_{ij}(x_i^{(2k+1)}+\omega_0^2 x_i^{(2k-1)}+i z_i^{(2k)}-\omega_0 z_i^{(2k-1)})\psi_j^{(2m+1)}+
\nonumber
\\[2pt]
&&
\qquad\qquad\;+\sum\limits_{k=0}^{n-1}\sum\limits_{m=0}^{n-k-1}\sigma_{k+m}^{n,1}\epsilon_{ij}(x_i^{(2k+2)}+\omega_0^2 x_i^{(2k)}-\omega_0 z_i^{(2k)})\psi_j^{(2m)}+
\\[2pt]
&&
\quad+i\sum\limits_{k=0}^{n-2}\sum\limits_{m=1}^{n-k-1}\sigma_{k+m}^{n,1}\epsilon_{ij}z_i^{(2k+1)}\psi_j^{(2m)}-
\epsilon_{ij}(\dot{x}_i-i\omega_0 x_i+iz_i)\sum\limits_{k=0}^{n-1}\sigma_k^{n,1}\psi_j^{(2k+1)},\quad \bar{Q}=Q^*.
\nonumber
\eea
These constants of motion, together with the Hamiltonian (\ref{H}), obey the following relations
\bea\label{alg}
[Q,Q\}=0,\qquad [H,Q\}=0,\qquad [Q,\bar{Q}\}=-2iH,\qquad [H,\bar{Q}\}=0,\qquad [\bar{Q},\bar{Q}\}=0,
\eea
with respect to the bracket (\ref{PB}). So, the model (\ref{action}) is an $\,\mathcal{N}=2$ supersymmetric extension of the odd-order PU oscillator (\ref{model}).

\vskip 0.5cm
\noindent
{\bf{\large 3. $\mathcal{N}=2$ supersymmetric third-order PU oscillator}}
\vskip 0.5cm
According to the analysis in Ref. \cite{Alt_1}, a Hamiltonian formulation of the odd-order PU oscillator (\ref{model}) is not unique. Let us generalize this result to the case of an $\,\mathcal{N}=2$ supersymmetric third-order PU oscillator. For $n=1$, the Hamiltonian of the model (\ref{model}) can be presented as a difference of two one-dimensional harmonic oscillators \cite{Alt_1}. This can be achieved by using the coordinates
\bea\label{qp}
\begin{aligned}
q_k=\frac{1}{\sqrt{2\omega_0}}\left(\dot{x}_1+\frac{(-1)^k}{\omega_0}\ddot{x}_2\right),\quad p_k=\sqrt{\frac{\omega_0}{2}}\left(\dot{x}_2+\frac{(-1)^{k+1}}{\omega_0}\ddot{x}_1\right),\quad y_k=\frac{1}{\omega_0}(\ddot{x}_k+\omega_0^2 x_k).
\end{aligned}
\eea
With respect to the supersymmetry transformations (\ref{2}), the variables $q_k$ and $y_k$ are transformed as follows
\bea
\d q_k=\vartheta_k\alpha+\bar{\vartheta}_k\bar{\alpha},\qquad
\d y_k=\theta_k\alpha+\bar{\theta}_k\bar{\alpha},
\nonumber
\eea
where we denoted
\bea\label{theta}
\vartheta_k=\frac{1}{\sqrt{2\omega_0}}\left(\dot{\psi}_1+i(-1)^k\dot{\psi}_2\right),\qquad \theta_k=i\dot{\psi}_k+\omega_0\psi_k, \qquad \bar{\vartheta}_k=(\vartheta_k)^*,\qquad \bar{\theta}_k=(\theta_k)^*.
\eea
The nonvanishing structure relations between the coordinates (\ref{qp}), (\ref{theta}) read
\bea
[q_k,p_m\}=\d_{km},\quad [y_k,y_m\}=-\epsilon_{km},\quad[\vartheta_k,\bar{\vartheta}_m\}=i(-1)^k\d_{km},\quad [\theta_k,\bar{\theta}_m\}=\omega_0\,\epsilon_{km},\quad\mbox{(no sum)}.
\nonumber
\eea

Using the variables (\ref{qp}), (\ref{theta}), the Hamiltonian (\ref{H}) and supercharges (\ref{Q}) for $n=1$ may be rewritten as\footnote{Note that the supersymmetry algebra (\ref{alg}) does not change when the supercharges are redefined as follows
$Q\rightarrow Q+\epsilon_{ij}\theta_i(y_j-z_j)$, $\bar{Q}\rightarrow \bar{Q}+\epsilon_{ij}\bar{\theta}_i(y_j-z_j)$.}
\bea
&&\label{12}
H=\frac{1}{2}(p_1^2+\omega_0^2q_1^2+2\omega_0\vartheta_1\bar{\vartheta}_1)-\frac{1}{2}(p_2^2+\omega_0^2 q_2^2+2\omega_0\vartheta_2\bar{\vartheta}_2),
\\[2pt]
&&\label{13}
Q=\vartheta_1(p_1-i\omega_0 q_1)+\vartheta_2(p_2+i\omega_0 q_2)-\epsilon_{ij}\theta_i(y_j-z_j),\qquad \bar{Q}=(Q)^*.
\eea
So, the Hamiltonian of an $\,\mathcal{N}=2$ supersymmetric third-order PU oscillator can be presented as a difference of two one-dimensional $\,\mathcal{N}=2$ supersymmetric harmonic oscillators. At first sight it may appear that an $\,\mathcal{N}=2$ supersymmetric odd-order PU oscillator is dynamically equivalent to a set of two decoupled $\,\mathcal{N}=2$ supersymmetric harmonic oscillators. This is not true because the phase spaces of these systems are not isomorphic. In addition to oscillator coordinates $(q_i,p_i,\vartheta_i,\bar{\vartheta}_i)$, the phase space of the $\,\mathcal{N}=2$ supersymmetric odd-order PU oscillator involves variables $a_i=\{y_i,z_i,\theta_i,\bar{\theta}_i\}$ whose dynamics are trivial $\dot{a}_i=0$. This also can be illustrated by rewriting the action functional (\ref{action}) as follows (up to a total derivative term)
\bea
&&
S=\frac{1}{2}\int dt\left[\left(\dot{q}_1^2-\omega_0^2 q_1^2+i\vartheta_1\dot{\bar{\vartheta}}_1+i\bar{\vartheta}_1\dot{\vartheta}_1-
2\omega_0\vartheta_1\bar{\vartheta}_1\right)+\epsilon_{ij}\left(y_i\dot{y}_j-z_i\dot{z}_j\right)-\right.
\nonumber
\\[2pt]
&&
\left.\qquad\qquad\;-\left(\dot{q}_2^2-\omega_0^2 q_2^2+i\vartheta_2\dot{\bar{\vartheta}}_2+i\bar{\vartheta}_2\dot{\vartheta}_2-
2\omega_0\vartheta_2\bar{\vartheta}_2\right)+\frac{1}{\omega_0}\epsilon_{ij}(\theta_i\dot{\bar{\theta}}_j-\bar{\theta}_i\dot{\theta}_j)\right].
\nonumber
\eea

Let us construct an alternative Hamiltonian formulation for an $\,\mathcal{N}=2$ supersymmetric third-order PU oscillator by applying the approach previously developed in Ref. \cite{Kosinski}. To this end, we must deform both the Hamiltonian (\ref{12}) and the corresponding Poisson bracket (\ref{PB}) in such a way that the equations (\ref{equ}) will be preserved. Let us choose the following deformation of the Hamiltonian (\ref{12})
\bea\label{Halt}
\mathcal{H}=\frac{\gamma_1}{2}(p_1^2+\omega_0^2q_1^2+2\omega_0\vartheta_1\bar{\vartheta}_1)+
\frac{\gamma_2}{2}(p_2^2+\omega_0^2 q_2^2+2\omega_0\vartheta_2\bar{\vartheta}_2),
\eea
where $\gamma_1$ and $\gamma_2$ are arbitrary nonzero coefficients.
With the change $H\rightarrow\mathcal{H}$, the equations (\ref{equ}) are satisfied provided the graded Poisson structure relations have the form
\bea\label{PSalt}
\begin{aligned}
&
[x_i,\ddot{x}_j\}=-\gamma^{-}\epsilon_{ij},&&[x_i,\dot{x}_j\}=\frac{1}{\omega_0}\gamma^{+}\d_{ij},&& [\psi_i,\dot{\bar{\psi}}_j\}=i\gamma^{-}\epsilon_{ij}-\gamma^{+}\d_{ij},\quad[z_i,z_j\}=\epsilon_{ij},
\\[2pt]
&
[\dot{x}_i,\dot{x}_j\}=\gamma^{-}\epsilon_{ij},&&[\dot{x}_i,\ddot{x}_j\}=\omega_0\gamma^{+}\d_{ij},&& [\dot{\psi}_i,\bar{\psi}_j\}=-i\gamma^{-}\epsilon_{ij}+\gamma^{+}\d_{ij},
\\[2pt]
&
[\ddot{x}_i,\ddot{x}_j\}=\omega_0^2\gamma^{-}\epsilon_{ij},&&[\psi_i,\bar{\psi}_j\}=-\frac{i}{\omega_0}\gamma^{+}\d_{ij},&& [\dot{\psi}_i,\dot{\bar{\psi}}_j\}=-\omega_0\gamma^{-}\epsilon_{ij}-i\omega_0\gamma^{+}\d_{ij},
\end{aligned}
\eea
where we denote
\bea
\gamma^{\pm}=\frac{1}{2}\left(\frac{1}{\gamma_1}\pm\frac{1}{\gamma_2}\right).
\nonumber
\eea
This Poisson structure is degenerate when $\gamma_1=\gamma_2$. By this reason, in what follows we exclude this case from our consideration.

Let us introduce the new variables
\bea\label{canon}
\begin{aligned}
&
\mathfrak{q}_k=\sqrt{|\gamma_k|}q_k, && \mathfrak{p}_k=(-1)^{k+1}\mbox{sign}(\gamma_k)\sqrt{|\gamma_k|}p_k, && \mathsf{y}_k=\frac{1}{\sqrt{|\gamma^{-}|}}y_k,
\\[2pt]
&
\Psi_k=\sqrt{|\gamma_k|}\vartheta_k,&& \bar{\Psi}_k=\sqrt{|\gamma_k|}\bar{\vartheta}_k,\quad \Theta_k=\frac{1}{\sqrt{|\gamma^{-}|}}\theta_k, && \bar{\Theta}_k=\frac{1}{\sqrt{|\gamma^{-}|}}\bar{\theta}_k.
\end{aligned}\quad\mbox{(no sum)}
\eea
Under the bracket (\ref{PSalt}), these coordinates obey the following nonvanishing relations
\bea\label{canon1}
\begin{aligned}
&
[\mathfrak{q}_k,\mathfrak{p}_m\}=\d_{km}, && [\mathsf{y}_k,\mathsf{y}_m\}=-\mbox{sign}(\gamma^{-})\epsilon_{km},
\\[4pt]
&
[\Psi_k,\bar{\Psi}_m\}=-i\,\mbox{sign}(\gamma_k)\d_{km}, && [\Theta_k,\bar{\Theta}_m\}=\omega_0\,\mbox{sign}(\gamma^-)\epsilon_{km},
\end{aligned}\quad\mbox{(no sum)}
\eea
Here and in what follows sign$(x)$ denotes the standard signum function. The Hamiltonian (\ref{Halt}) in terms of the variables (\ref{canon}) takes the form
\bea\label{Halt1}
\mathcal{H}=\frac{\mbox{sign}(\gamma_1)}{2}(\mathfrak{p}_1^2+\omega_0^2 \mathfrak{q}_1^2+2\omega_0\Psi_1\bar{\Psi}_1)+
\frac{\mbox{sign}(\gamma_2)}{2}(\mathfrak{p}_2^2+\omega_0^2 \mathfrak{q}_2^2+2\omega_0\Psi_2\bar{\Psi}_2).
\eea

Along with this alternative Hamiltonian, the full formulation of an $\,\mathcal{N}=2$ supersymmetric third-order PU oscillator involves supercharges.
According to the analysis in Ref. \cite{Alt}, one may try to find these by using an auxiliary action functional. Taking into account the relations (\ref{canon1}), in our case such an action can be chosen in the form
\bea
\begin{aligned}
&
\mathcal{S}=\frac{1}{2}\int dt\,\mbox{sign}(\gamma_1)(\dot{\mathfrak{q}}_1^2-\omega_0^2\mathfrak{q}_1^2+i\Psi_1\dot{\bar{\Psi}}_1+
i\bar{\Psi}_1\dot{\Psi}_1-2\omega_0\Psi_1\bar{\Psi}_1)+\mbox{sign}(\gamma^{-})\epsilon_{ij}\mathsf{y}_i\dot{\mathsf{y}_j}-\epsilon_{ij}z_i\dot{z}_j+
\\[2pt]
&
\qquad+\mbox{sign}(\gamma_2)(\dot{\mathfrak{q}}_2^2-\omega_0^2\mathfrak{q}_2^2+i\Psi_2\dot{\bar{\Psi}}_2+
i\bar{\Psi}_2\dot{\Psi}_2-2\omega_0\Psi_2\bar{\Psi}_2)+
\frac{\mbox{sign}(\gamma^{-})}{\omega_0}\epsilon_{ij}\left(\Theta_i\dot{\bar{\Theta}}_j-\bar{\Theta}_i\dot{\Theta}_j\right).
\end{aligned}
\nonumber
\eea
This action is invariant under the transformations
\bea
\begin{aligned}
&
\d\mathfrak{q}_k=\Psi_k\alpha+\bar{\Psi}_k\bar{\alpha}, && \d\Psi_k=(-i\dot{\mathfrak{q}}_k+\omega_0\mathfrak{q}_k)\bar{\alpha}, &&
\d\Theta_k=\omega_0(\mathsf{y}_k-\mbox{sign}(\gamma^{-})z_k)\bar{\alpha},
\\[2pt]
&
\d\mathsf{y}_k=\d z_k=\Theta_k\alpha+\bar{\Theta}_k\bar{\alpha}, && \d\bar{\Psi}_k=(-i\dot{\mathfrak{q}}_k-\omega_0\mathfrak{q}_k)\alpha, && \d\bar{\Theta}_k=-\omega_0(\mathsf{y}_k-\mbox{sign}(\gamma^{-})z_k)\alpha,
\end{aligned}
\nonumber
\eea
which yield the following Noether integrals of motion
\bea
\mathcal{Q}=\Psi_1(\mathfrak{p}_1-i\,\mbox{sign}(\gamma_1)\omega_0\mathfrak{q}_1)+\Psi_2(\mathfrak{p}_2-i\,\mbox{sign}(\gamma_2)\omega_0\mathfrak{q}_2)
-\epsilon_{ij}\Theta_i(\mbox{sign}(\gamma^{-})\mathsf{y}_j-z_j),\;\bar{\mathcal{Q}}=(\mathcal{Q})^*.
\nonumber
\eea
With respect to the alternative Poisson structure (\ref{PSalt}), these conserved quantities and the alternative Hamiltonian (\ref{Halt1}) obey the relations
\bea
\begin{aligned}
&
 [\mathcal{H},\mathcal{Q}\}=0,\qquad [\mathcal{Q},\bar{\mathcal{Q}}\}=-2i\mathcal{H}+(1-\mbox{sign}(\gamma^{-}))\epsilon_{ij}\Theta_i\bar{\Theta}_j,\qquad [\mathcal{H},\bar{\mathcal{Q}}\}=0,
\\[4pt]
&
\qquad[\mathcal{Q},\mathcal{Q}\}=(1-\mbox{sign}(\gamma^{-}))\epsilon_{ij}\Theta_i\Theta_j,\quad\; [\bar{\mathcal{Q}},\bar{\mathcal{Q}}\}=(1-\mbox{sign}(\gamma^{-}))\epsilon_{ij}\bar{\Theta}_i\bar{\Theta}_j,
\end{aligned}
\nonumber
\eea
Thus, for positive $\gamma^{-}$, we have an appropriate supercharges $\mathcal{Q}$ and $\bar{\mathcal{Q}}$.
Moreover, if we put $0<\gamma_1<\gamma_2$ then the corresponding alternative Hamiltonian becomes a direct sum of two one-dimensional $\,\mathcal{N}=2$ supersymmetric harmonic oscillators.

\vskip 0.5cm
\noindent
{\bf{\large 4. The general case}}
\vskip 0.5cm

Let us consider an $\,\mathcal{N}=2$ supersymmetric PU oscillator of arbitrary odd order. To construct an alternative Hamiltonian formulation for this system, one should obtain a more appropriate representation for the Hamiltonian (\ref{H}). According to the analysis in Ref. \cite{Alt_1}, a Hamiltonian of the model (\ref{model}) can be represented as a direct sum of the third-order PU oscillators which alternate in a sign. This can be achieved with the aid of the so-called oscillator variables \cite{Pais,Alt_1}
\bea\label{oc}
\tilde{x}_{k,i}=\sqrt{\rho_k^{n,0}}\prod_{m=0\atop m\neq k}^{n-1}\left(\frac{d^2}{dt^2}+\omega_m^2\right)\dot{x}_i,\quad z_{0,i}=\frac{1}{\prod\limits_{s=0}^{n-1}\omega_s}\prod_{m=0}^{n-1}\left(\frac{d^2}{dt^2}+\omega_m^2\right)x_i,
\eea
where $k=0,1,..,n-1$; the coefficients $\rho_k^{n,s}$ are given by
\bea
\rho^{n,s}_{k}=\frac{(-1)^{k+s}}{\prod\limits_{m=s\atop m\neq k}^{n-1}(\omega_m^2-\omega_k^2)},\qquad k=s,s+1,..,n-1.
\nonumber
\eea
Taking into account the results of Refs. \cite{Alt,Masterov_N2PU}, let us introduce similar variables for the remaining coordinates
\bea\label{oc1}
\begin{aligned}
&
\psi_{p,i}=\sqrt{\rho_p^n}\prod\limits_{m=-n+1\atop m\neq p}^{n-1}\left(\frac{d}{dt}-i\omega_m\right)\dot{\psi}_i, && \theta_i=\frac{i}{\prod\limits_{s=1}^{n-1}\omega_s}\prod_{m=-n+1}^{n-1}\left(\frac{d}{dt}-i\omega_m\right)\psi_i, &&
\bar{\psi}_{p,i}=(\psi_{p,i})^*,
\\[2pt]
&
\tilde{x}_{-k,i}=\sqrt{\rho_k^{n,1}}\prod_{m=1\atop m\neq k}^{n-1}\left(\frac{d^2}{dt^2}+\omega_m^2\right)\dot{z}_i, &&
z_{1,i}=\frac{1}{\prod\limits_{s=1}^{n-1}\omega_s}\prod_{m=1}^{n-1}\left(\frac{d^2}{dt^2}+\omega_m^2\right)z_i, &&
\bar{\theta}_i=(\theta_i)^*,
\end{aligned}
\eea
where $k=1,2,..,n-1$, $p=-n+1,-n+2,..,n-1$; the coefficients $\rho_p^n$ are defined by
\bea
\rho_p^n=\frac{(-1)^{n+p-1}}{\prod\limits_{m=-n+1\atop m\neq p}^{n-1}(\omega_m-\omega_p)}=\frac{\omega_p+\omega_0}{2\omega_p}\rho_{|p|}^{n,0}.
\nonumber
\eea

Let us draw our attention to how the variables $\tilde{x}_{\pm k,i}$, $\psi_{\pm k,i}$, and $\bar{\psi}_{\pm k,i}$ ($k=1,2,..,n-1$) are transformed under the supersymmetry transformations (\ref{2})
\bea\label{susy}
\begin{aligned}
&
\d \tilde{x}_{\pm k,i}
=(\mu_k^{\pm}\psi_{k,i}\pm\mu_{k}^{\mp}\psi_{-k,i})\alpha+(\mu_k^{\pm}\bar{\psi}_{k,i}\pm\mu_k^{\mp}\bar{\psi}_{-k,i})\bar{\alpha},
\\[2pt]
&
\d\psi_{\pm k,i}=(-\mu_k^{\pm}(i\dot{\tilde{x}}_{k,i}\mp\omega_k \tilde{x}_{k,i})\mp\mu_k^{\mp}(i\dot{\tilde{x}}_{-k,i}\mp\omega_k \tilde{x}_{-k,i}))\bar{\alpha},
\\[2pt]
&
\d\bar{\psi}_{\pm k,i}=(-\mu_k^{\pm}(i\dot{\tilde{x}}_{k,i}\pm\omega_k\tilde{x}_{k,i})\mp\mu_k^{\mp}(i\dot{\tilde{x}}_{-k,i}\pm\omega_k \tilde{x}_{-k,i}))\alpha,
\end{aligned}\quad\mbox{with}\quad\mu_{k}^{\pm}=\sqrt{\frac{\omega_k\pm\omega_0}{2\omega_k}}.
\eea
This motivates us to perform one more change of the bosonic coordinates
\bea\label{x}
x_{-k,i}=\mu_k^{-}\tilde{x}_{k,i}-\mu_k^{+}\tilde{x}_{-k,i},\quad x_{0,i}=\tilde{x}_{0,i},\quad x_{k,i}=\mu_k^{+}\tilde{x}_{k,i}+\mu_k^{-}\tilde{x}_{-k,i}.
\eea
The supersymmetry transformations (\ref{susy}) then become
\bea
\d x_{\pm k,i}=\psi_{\pm k,i}\alpha+\bar{\psi}_{\pm k,i}\bar{\alpha},\quad \d\psi_{\pm k,i}=-(i\dot{x}_{\pm k,i}\mp\omega_k x_{\pm k,i})\bar{\alpha},\quad \d\bar{\psi}_{\pm k,i}=-(i\dot{x}_{\pm k,i}\pm\omega_k x_{\pm k,i})\alpha.
\nonumber
\eea

The Hamiltonian (\ref{H}) and the supercharges (\ref{Q}) in terms of $x_{k,i}$, $\psi_{k,i}$, $\bar{\psi}_{k,i}$, and $z_{s,i}$ may be represented as follows
\bea
&&
H=\sum_{k=-n+1}^{n-1}(-1)^{k+1}\epsilon_{ij}(x_{k,i}\dot{x}_{k,j}-i\psi_{k,i}\bar{\psi}_{k,j}),
\nonumber
\\[2pt]
&&
Q=\sum_{k=-n+1}^{n-1}\frac{(-1)^k}{\omega_k}\epsilon_{ij}\psi_{k,i}(i\dot{x}_{k,j}+\omega_k x_{k,j})-\epsilon_{ij}\theta_i(z_{0,j}-z_{1,j}),\quad \bar{Q}=(Q)^*.
\nonumber
\eea
So, we have shown that the Hamiltonian of an $\,\mathcal{N}=2$ supersymmetric $(2n+1)$-order PU oscillator can be presented as a direct sum of $(2n-1)$ $\,\mathcal{N}=2$ supersymmetric third-order PU oscillators which alternate in a sign. This fact correlates with the analysis in Ref. \cite{Alt_1} for the model (\ref{model}).

By analogy with (\ref{qp}), (\ref{theta}), let us introduce the coordinates
\bea\label{var}
\begin{aligned}
&
q_{k,s}=\frac{1}{\sqrt{|2\omega_k|}}\left(x_{k,1}+\frac{(-1)^s}{|\omega_k|}\dot{x}_{k,2}\right), &&
p_{k,s}=(-1)^k\sqrt{\frac{|\omega_k|}{2}}\left(x_{k,2}+\frac{(-1)^{s+1}}{|\omega_k|}\dot{x}_{k,1}\right),
\\[2pt]
&
\vartheta_{k,s}=\frac{1}{\sqrt{|2\omega_k|}}(\psi_{k,1}+i(-1)^s\,\mbox{sign}(\omega_k)\psi_{k,2}), &&
\bar{\vartheta}_{k,s}=\frac{1}{\sqrt{|2\omega_k|}}(\bar{\psi}_{k,1}-i(-1)^s\,\mbox{sign}(\omega_k)\bar{\psi}_{k,2}).
\end{aligned}
\eea
Given the bracket (\ref{PB}), these variables obey
\bea
[q_{k,s},p_{m,j}\}=\d_{km}\d_{sj},\qquad [\vartheta_{k,s},\bar{\vartheta}_{m,j}\}=i(-1)^{k+s}\d_{km}\d_{sj}.\qquad\mbox{(no sum)}
\nonumber
\eea
The existence of these coordinates automatically implies that (\ref{PB}) possesses standard properties of a graded Poisson bracket.

In terms of the variables (\ref{var}), the Hamiltonian (\ref{H}) takes the form
\bea
&&
H=\sum_{k=-n+1}^{n-1}(-1)^k\left[\left(\frac{1}{2}p_{k,1}^2+\frac{\omega_k^2}{2}q_{k,1}^2+\omega_k\vartheta_{k,1}\bar{\vartheta}_{k,1}\right)-
\left(\frac{1}{2}p_{k,2}^2+\frac{\omega_k^2}{2}q_{k,2}^2+\omega_k\vartheta_{k,2}\bar{\vartheta}_{k,2}\right)\right].
\nonumber
\eea
Let us consider the following deformation of this Hamiltonian
\bea\label{Hdef}
\begin{aligned}
&
\mathcal{H}=\sum_{k=-n+1}^{n-1}\gamma_{|k|,1}\left(\frac{1}{2}p_{k,1}^2+\frac{\omega_k^2}{2} q_{k,1}^2+\omega_{k}\vartheta_{k,1}\bar{\vartheta}_{k,1}\right)+\gamma_{|k|,2}
\left(\frac{1}{2}p_{k,2}^2+\frac{\omega_k^2}{2} q_{k,2}^2+\omega_{k}\vartheta_{k,2}\bar{\vartheta}_{k,2}\right),
\end{aligned}
\eea
where $\gamma_{0,1},\gamma_{0,2},\gamma_{1,1},..,\gamma_{n-1,2}$ are arbitrary nonzero coefficients. It is straightforward to verify (for technical details see Appendix) that the equations (\ref{equ}), where $H\rightarrow\mathcal{H}$, are satisfied provided the following graded Poisson structure
\bea\label{Table}
\begin{array}{|c|c|c|}
\hline
& [x_i^{(s)},x_j^{(m)}\} & [z_i^{(s)},z_j^{(m)}\} \\
\hline
s=m=0 & 0 & 0\\[2pt]
s+m-odd & (-1)^{\frac{s-m+1}{2}}\sum\limits_{k=0}^{n-1}\rho_k^{n,0}\omega_k^{s+m-2}\gamma_k^{+}\d_{ij} &  (-1)^{\frac{s-m+1}{2}}\sum\limits_{k=1}^{n-1}\rho_k^{n,1}\omega_k^{s+m-2}\gamma_k^{+}\d_{ij} \\[2pt]
s+m\neq0-even & (-1)^{\frac{s-m}{2}}\sum\limits_{k=0}^{n-1}\rho_k^{n,0}\omega_k^{s+m-2}\gamma_k^{-}\epsilon_{ij} & (-1)^{\frac{s-m}{2}}\sum\limits_{k=1}^{n-1}\rho_k^{n,1}\omega_k^{s+m-2}\gamma_k^{-}\epsilon_{ij}\\
\hline
\multicolumn{1}{|c|}{} & \multicolumn{2}{c|}{[\psi_i^{(s)},\bar{\psi}_j^{(m)}\}}\\
\hline
\multicolumn{1}{|c|}{s=m=0}&\multicolumn{2}{c|}{-i\sum\limits_{k=0}^{n-1}\rho_k^{n,0}\omega_k^{-1}\gamma_k^{+}\d_{ij}}\\[7pt]
\multicolumn{1}{|c|}{s+m-odd}&\multicolumn{2}{c|}{(-1)^{\frac{s-m-1}{2}}\left(\omega_0\sum\limits_{k=0}^{n-1}\rho_k^{n,0}\omega_k^{s+m-2}\gamma_k^{+}\d_{ij}-
i\sum\limits_{k=0}^{n-1}\rho_k^{n,0}\omega_k^{s+m-1}\gamma_k^{-}\epsilon_{ij}\right)}\\[7pt]
\multicolumn{1}{|c|}{s+m\neq 0-even}&\multicolumn{2}{c|}{(-1)^{\frac{s-m-2}{2}}\left(i\sum\limits_{k=0}^{n-1}\rho_k^{n,0}\omega_k^{s+m-1}\gamma_k^{+}\d_{ij}+
\omega_0\sum\limits_{k=0}^{n-1}\rho_k^{n,0}\omega_k^{s+m-2}\gamma_k^{-}\epsilon_{ij}\right)}\\
\hline
\end{array}
\eea
has been chosen.
Above we denote $\gamma_{k}^{\pm}=\frac{1}{2}\left(\frac{1}{\gamma_{k,1}}\pm\frac{1}{\gamma_{k,2}}\right)$. This structure is degenerate provided $g_{n,0}=0$ and/or $g_{n,1}=0$, where
\bea
g_{n,s}=\sum_{k=s}^{n-1}\frac{\rho_k^{n,s}}{2\omega_k^2}\left(\frac{1}{\gamma_{k,1}}-\frac{1}{\gamma_{k,2}}\right)=
\sum_{k=s}^{n-1}\frac{\rho_k^{n,s}\gamma_{k}^{-}}{\omega_k^2}.
\nonumber
\eea
By this reason, we restrict our consideration only to the case when $g_{n,0}\neq 0$, $g_{n,1}\neq 0$.

The generalization of the coordinates (\ref{canon}) reads
\bea
\begin{aligned}
&
\mathfrak{q}_{k,i}=\sqrt{|\gamma_{|k|,i}|}q_{k,i}, && \mathsf{z}_{s,i}=\frac{1}{\prod\limits_{m=s}^{n-1}\omega_m\sqrt{|g_{n,s}|}}z_{s,i}, && \mathfrak{p}_{k,i}=(-1)^{k+i+1}\mbox{sign}(\gamma_{k,i})\sqrt{|\gamma_{|k|,i}|}p_{k,i},
\\[2pt]
&
\Psi_{k,i}=\sqrt{|\gamma_{|k|,i}|}\vartheta_{k,i},&& \Theta_i=\frac{1}{\prod\limits_{m=0}^{n-1}\omega_m\sqrt{|g_{n,0}|}}\theta_i,&&
\bar{\Psi}_{k,i}=(\Psi_{k,i})^*,\; \bar{\Theta}_i=(\Theta_i)^*.\;\mbox{(no sum)}
\end{aligned}
\eea
With respect to the Poisson structure (\ref{Table}), these variables obey the relations
\bea\label{GPB}
\begin{aligned}
&
[\mathfrak{q}_{k,i},\mathfrak{p}_{m,j}\}=\d_{km}\d_{ij}, && [\mathsf{z}_{s,i},\mathsf{z}_{m,j}\}=-\mbox{sign}(g_{n,s})\d_{sm}\epsilon_{ij},
\\[2pt]
&
[\Theta_i,\bar{\Theta}_j\}=\omega_0\,\mbox{sign}(g_{n,0})\epsilon_{ij}, &&
[\Psi_{k,i},\bar{\Psi}_{m,j}\}=-i\,\mbox{sign}(\gamma_{|k|,i})\d_{km}\d_{ij}.\;\mbox{(no sum)}
\end{aligned}
\eea
Then the alternative Hamiltonian (\ref{Hdef}) may be rewritten as
\bea\label{HH}
\mathcal{H}=\sum_{k=-n+1}^{n-1}\frac{\mbox{sign}(\gamma_{|k|,i})}{2}\left(\mathfrak{p}_{k,i}^2+\omega_k^2 \mathfrak{q}_{k,i}^2+2\omega_k\Psi_{k,i}\bar{\Psi}_{k,i}\right).
\eea

To find supercharges corresponding to this alternative Hamiltonian, let us introduce the following auxiliary action functional
\bea
&&
\mathcal{S}=\frac{1}{2}\int dt\sum_{k=-n+1}^{n-1}\mbox{sign}(\gamma_{|k|,i})(\dot{\mathfrak{q}}_{k,i}^2-\omega_k^2\mathfrak{q}_{k,i}^2+i\Psi_{k,i}\dot{\bar{\Psi}}_{k,i}+
i\bar{\Psi}_{k,i}\dot{\Psi}_{k,i}-2\omega_k\Psi_{k,i}\bar{\Psi}_{k,i})+
\nonumber
\\[2pt]
&&
\qquad\qquad\qquad+\sum_{s=0}^{1}\mbox{sign}(g_{n,s})\epsilon_{ij}\mathsf{z}_{s,i}\dot{\mathsf{z}}_{s,j}
+\frac{\mbox{sign}(g_{n,0})}{\omega_0}\epsilon_{ij}\left(\Theta_i\dot{\bar{\Theta}}_j-\bar{\Theta}_i\dot{\Theta}_j\right),
\nonumber
\eea
which is invariant under the transformations
\bea
\begin{aligned}
&
\d\mathfrak{q}_{k,i}=\Psi_{k,i}\alpha+\bar{\Psi}_{k,i}\bar{\alpha},&& \d\Psi_{k,i}=(-i\dot{\mathfrak{q}}_{k,i}+\omega_k\mathfrak{q}_{k,i})\bar{\alpha},&& \d\Theta_i=\omega_0(\mathsf{z}_{0,i}+\mbox{sign}(g_{n,0}g_{n,1})\mathsf{z}_{1,i})\bar{\alpha},
\\[2pt]
&
\d\mathsf{z}_{k,i}=\Theta_i\alpha+\bar{\Theta}_i\bar{\alpha},&&\d\bar{\Psi}_{k,i}=(-i\dot{\mathfrak{q}}_{k,i}-\omega_k\mathfrak{q}_{k,i})\alpha,&& \d\bar{\Theta}_i=-\omega_0(\mathsf{z}_{0,i}+\mbox{sign}(g_{n,0}g_{n,1})\mathsf{z}_{1,i})\alpha.
\end{aligned}
\nonumber
\eea
The Noether charges associated with these symmetries read
\bea
\begin{aligned}
&&
\mathcal{Q}=\sum_{k=-n+1}^{n-1}\Psi_{k,i}(\mathfrak{p}_{k,i}-i\,\mbox{sign}(\gamma_{|k|,i})\omega_k\mathfrak{q}_{k,i})-
\epsilon_{ij}\Theta_i(\mbox{sign}(g_{n,0})\mathsf{z}_{0,j}+\mbox{sign}(g_{n,1})\mathsf{z}_{1,j}),\;\bar{\mathcal{Q}}=(\mathcal{Q})^*.
\end{aligned}
\nonumber
\eea
These integrals of motion, together with the Hamiltonian (\ref{HH}), obey the following relations
\bea
\begin{aligned}
&
[\mathcal{H},\mathcal{Q}\}=0,\qquad\;\, [\mathcal{Q},\bar{\mathcal{Q}}\}=-2i\mathcal{H}-
(\mbox{sign}(g_{n,0})+\mbox{sign}(g_{n,1}))\epsilon_{ij}\Theta_i\bar{\Theta}_j,\qquad\;\, [\mathcal{H},\bar{\mathcal{Q}}\}=0,
\\[4pt]
&
[\mathcal{Q},\mathcal{Q}\}=-(\mbox{sign}(g_{n,0})+\mbox{sign}(g_{n,1}))\epsilon_{ij}\Theta_i\Theta_j,\quad\; [\bar{\mathcal{Q}},\bar{\mathcal{Q}}\}=-(\mbox{sign}(g_{n,0})+\mbox{sign}(g_{n,1}))\epsilon_{ij}\bar{\Theta}_i\bar{\Theta}_j,
\end{aligned}
\nonumber
\eea
under the bracket (\ref{GPB}). Thus, we have one more condition on the coefficients $\gamma_{k,i}$
\bea\label{sign}
\mbox{sign}(g_{n,0})=-\mbox{sign}(g_{n,1}).
\eea
It is evident that infinitely many possible sets of parameters $\gamma_{k,i}$ obey this restriction.

\vskip 0.5cm
\noindent
{\bf{\large 5. Quantization}}
\vskip 0.5cm

To quantize an $\,\mathcal{N}=2$ supersymmetric odd-order PU oscillator with the Hamiltonian (\ref{HH}), let us introduce hermitian bosonic operators $\hat{\mathfrak{q}}_{k,i}$, $\hat{\mathfrak{p}}_{k,i}$ $\hat{\mathsf{z}}_{s,i}$ as well as fermionic operators $\hat{\Psi}_{k,i}$, $\hat{\bar{\Psi}}_{k,i}=(\hat{\Psi}_{k,i})^\dagger$, $\hat{\Theta}_i$, $\hat{\bar{\Theta}}_i=(\hat{\Theta}_i)^\dagger$. According to (\ref{GPB}) and (\ref{sign}), they obey the following nonvanishing (anti)commutation relations
\bea\label{ACR}
\begin{aligned}
&
[\hat{\mathfrak{q}}_{k,i},\hat{\mathfrak{p}}_{m,j}]=i\hbar\d_{km}\d_{ij}, && [\hat{\mathsf{z}}_{s,i},\hat{\mathsf{z}}_{m,j}]=-i(-1)^s\hbar\,\mbox{sign}(g_{n,0})\d_{sm}\epsilon_{ij},
\\[2pt]
&
\{\hat{\Theta}_i,\hat{\bar{\Theta}}_j\}=i\hbar\omega_0\,\mbox{sign}(g_{n,0})\epsilon_{ij}, &&
\{\hat{\Psi}_{k,i},\hat{\bar{\Psi}}_{m,j}\}=\hbar\,\mbox{sign}(\gamma_{|k|,i})\d_{km}\d_{ij},\;\mbox{(no sum)}
\end{aligned}
\eea
where $\{\cdot,\cdot\}$ and $[\cdot,\cdot]$ stand for the anticommutator and commutator, respectively. $\hbar$ is the reduced Planck constant.

As the next step, we may introduce the creation $\bar{a}_{k,i}$, $\bar{c}_{k,i}$ and annihilation $a_{k,i}$, $c_{k,i}$ operators, which correspond to oscillator coordinates $(\mathfrak{q}_{k,i},\mathfrak{p}_{k,i},\Psi_{k,i},\bar{\Psi}_{k,i})$
\bea
\begin{aligned}
&
a_{k,i}=\sqrt{\frac{|\omega_k|}{2\hbar}}\hat{\mathfrak{q}}_{k,i}+i\frac{1}{\sqrt{2|\omega_k|\hbar}}\hat{\mathfrak{p}}_{k,i},\qquad c_{k,i}=\frac{1}{\sqrt{\hbar}}\hat{\Psi}_{k,i},
\\[2pt]
&
\bar{a}_{k,i}=\sqrt{\frac{|\omega_k|}{2\hbar}}\hat{\mathfrak{q}}_{k,i}-i\frac{1}{\sqrt{2|\omega_k|\hbar}}\hat{\mathfrak{p}}_{k,i},\qquad \bar{c}_{k,i}=\frac{1}{\sqrt{\hbar}}\hat{\bar{\Psi}}_{k,i},
\end{aligned}\;\Rightarrow\;
\begin{aligned}
&
[a_{k,i},\bar{a}_{m,j}]=\d_{km}\d_{ij},
\\[2pt]
&
\{c_{k,i},\bar{c}_{m,j}\}=\mbox{sign}(\gamma_{|k|,i})\d_{km}\d_{ij}.
\end{aligned}
\nonumber
\eea
Thus, for negative values of $\gamma_{k,i}$, we have $\{c_{k,i},\bar{c}_{m,j}\}=-\d_{km}\d_{ij}$. Taking into account the analysis in Refs. \cite{Masterov_N2PU,MP}, these relations bring about negative norm states. To avoid this feature, we set all coefficients $\gamma_{k,i}$ to be positive.

For the variables $\mathsf{z}_{s,i}$, $\Theta_i$, and $\bar{\Theta}_i$, the creation $\bar{b}_s$, $\bar{d}_s$ and annihilation $b_s$, $d_s$ operators may be defined as follows \cite{Lozano}
\bea
\begin{aligned}
&
b_s=\frac{1}{\sqrt{2\hbar}}(\hat{\mathsf{z}}_{s,1}-i(-1)^s\mbox{sign}(g_{n,0})\hat{\mathsf{z}}_{s,2}), &&
d_s=\frac{1}{\sqrt{2\hbar\omega_0}}(\hat{\Theta}_1+i(-1)^s\mbox{sign}(g_{n,0})\hat{\Theta}_2),
\\[2pt]
&
\bar{b}_s=\frac{1}{\sqrt{2\hbar}}(\hat{\mathsf{z}}_{s,1}+i(-1)^s\mbox{sign}(g_{n,0})\hat{\mathsf{z}}_{s,2}), &&
\bar{d}_s=\frac{1}{\sqrt{2\hbar\omega_0}}(\hat{\bar{\Theta}}_1-i(-1)^s\mbox{sign}(g_{n,0})\hat{\bar{\Theta}}_2),
\end{aligned}\;s=0,1.
\nonumber
\eea
These operators obey the following nonvanishing relations
\bea
[b_s,\bar{b}_p]=\d_{sp},\qquad \{d_s,\bar{d}_p\}=(-1)^s\d_{sp}.
\nonumber
\eea
Unfortunately, the relation $\{d_1,\bar{d}_1\}=-1$ leads to the presence of negative norm states in the corresponding Fock space \cite{Masterov_N2PU,MP}.

\vskip 0.5cm
\noindent
{\bf{\large 6. Conclusion}}
\vskip 0.5cm
To summarize, in this work we have introduced an $\,\mathcal{N}=2$ supersymmetric generalization for the odd-order PU oscillator with distinct frequencies of oscillation. This system is invariant under the time translations. We have observed that the corresponding integral of motion can be presented as a direct sum of one-dimensional $\,\mathcal{N}=2$ supersymmetric harmonic oscillators which alternate in a sign. This representation has allowed us to construct a family of Hamiltonian structures for an $\,\mathcal{N}=2$ supersymmetric odd-order PU oscillator. Unfortunately, quantization of the system revealed the presence of negative norm states in the corresponding Fock space.

Turning to further possible developments, it is worth constructing various generalizations of an $\,\mathcal{N}=2$ supersymmetric odd-order PU oscillator which are compatible with the alternative Hamiltonian formulation. In particular, it would be interesting to generalize deformed odd-order PU oscillator introduced in paper \cite{Alt_1} as well as higher-derivative field theories considered in \cite{Dima_2} to an $\,\mathcal{N}=2$ supersymmetric case. A construction of $\,\mathcal{N}=2$ supersymmetric many particle higher-derivative systems is also of interest. In this context it is worth studying higher-derivative generalizations of $\,\mathcal{N}=2$ supersymmetric many body models constructed in papers \cite{N=2mb}-\cite{Galajinsky_4}.
The odd-order PU oscillator with weak supersymmetry \cite{weak} has been introduced in paper \cite{Masterov_3}. It is also worth investigating a Hamiltonian formulation of this system. These issues will be studied elsewhere.

\vskip 0.5cm
\noindent
{\large{\bf Acknowledgements}}
\vskip 0.5cm
We would like to thank D. Chow for pointing out reference \cite{idty_1}. We also thank an anonymous Referee for useful comments.
This work was supported by the MSE program Nauka under the project 3.825.2014/K, and RFBR grant 15-52-05022.

\vskip 0.5cm
\noindent
{\large{\bf Appendix:} List of identities}
\vskip 0.5cm
When verifying the fact that the equations (\ref{equ}) are satisfied with respect to the Hamiltonian structures introduced in both Sects. 2 and 4, the following identities
\bea
\begin{aligned}
&
P_{2k}^{n,s}=\sum_{p=s}^{n-1}(-1)^{n+p+1}\omega_p^{2n+2k-2s-2}\rho_p^{n,s},\quad k=-n+s+1,-n+s+2,..;
\nonumber
\\[2pt]
&
\sum_{k=s}^{n-1}(-1)^{k+s}(-\omega_k^2)^{r}\sigma_{p,k}^{n,s}\rho_k^{n,s}=
\left\{
\begin{aligned}
&
\d_{rp},&& r=0,1,..,n-s-1;
\\[5pt]
&
-\sigma_p^{n,s},&& r=n-s;
\end{aligned}
\right.\quad \sum_{r=0}^{n-1}\frac{(-1)^r}{\omega_r^2}\sigma_{p,r}^{n,0}\rho_r^{n,0}=\frac{\sigma_{p+1}^{n,0}}{\prod\limits_{k=0}^{n-1}\omega_k^2};
\nonumber
\\[2pt]
&
\sum_{r=0}^{n-s-1}(-1)^r\omega_q^{2r}\sigma_{r,k}^{n,s}=\frac{(-1)^{k+s}}{\rho_k^{n,s}}\d_{qk};\quad \sigma_{p,k}^{n,s}=\sum_{r=0}^{n-p-s-1}(-1)^r\omega_k^{2r}\sigma_{p+r+1}^{n,s},\quad k=s,s+1,..,n-1;
\end{aligned}
\nonumber
\eea
with
\bea
\sigma_{m,k}^{n,s}=\sum\limits_{i_1<i_2<..<i_{n-m-1}=s\atop i_1,i_2,..,i_{n-m-1}\neq k}^{n-1}\omega_{i_1}^2\omega_{i_2}^2..\omega_{i_{n-m-1}}^2,\quad \sigma_{n-s-1,k}^{n,s}\equiv 1,
\nonumber
\eea
prove to be helpful. The proofs of these identities can be found in Refs. \cite{Alt,Alt_1}, \cite{idty_1}.

\fontsize{10}{13}\selectfont


\begin{thebibliography}{nn}
\bibitem{Ostrogradski}
M.V. Ostrogradski,\\ {\it Memoires sur les equations differentielles relatives au probleme des isoperimetretes}, \\Mem. Ac. St.-Petersbourg {\bf VI 4} (1850) 385-517.
\bibitem{Dirac}
P.A.M. Dirac,\\ {\it Lectures on Quantum Mechanics},\\ Belfer Graduate School of Science, Yeshiva University (1964).
\bibitem{Faddeev}
L. Faddeev, R. Jackiw,\\ {\it Hamiltonian reduction of unconstrained and constrained systems},\\ Phys. Rev. Lett. {\bf 60} (1988) 1692-1694.
\bibitem{Pais}
A. Pais, G.E. Uhlenbeck,\\ {\it On field theories with nonlocalized action},\\ Phys. Rev. {\bf 79} (1950) 145-165.
\bibitem{Woodard}
R.P. Woodard,\\ {\it Avoiding dark energy with 1/r modifications of gravity},\\ Lect. Notes Phys. {\bf 720} (2007) 403-433, [astro-ph/0601672].
\bibitem{Smilga}
A.V. Smilga,\\ {\it Comments on the dynamics of the Pais-Uhlenbeck oscillator},\\ SIGMA {\bf 5} (2009) 017, arXiv:0808.0139 [quant-ph].
\bibitem{Kosinski}
K. Bolonek, P. Kosi\'{n}ski,\\ {\it Hamiltonian structures for Pais-Uhlenbeck oscillator},\\ Acta Phys. Polon. B {\bf 36} (2005) 2115-2131, [quant-ph/0501024].
\bibitem{KLS}
D.S. Kaparulin, S.L. Lyakhovich, A.A. Sharapov,\\ {\it Classical and quantum stability of higher-derivative dynamics},\\ Eur. Phys. J. C {\bf 74} (2014) 10, 3072, arXiv:1407.8481 [hep-th].
\bibitem{Ali_1}
A. Mostafazadeh,\\ {\it A Hamiltonian formulation of the Pais-Uhlenbeck oscillator that yields a stable and unitary quantum system}, Phys. Lett. A {\bf 375} (2010) 93-98, arXiv:1008.4678 [hep-th].
\bibitem{Alt}
I. Masterov,\\ {\it An alternative Hamiltonian formulation for the Pais-Uhlenbeck oscillator},\\ Nucl. Phys. B {\bf 902} (2016) 95-114, arXiv:1505.02583 [hep-th].
\bibitem{Alt_1}
I. Masterov,\\ {\it The odd-order Pais-Uhlenbeck oscillator},\\ Nucl. Phys. B {\bf 907} (2016) 495-508, arXiv:1603.07727 [math-ph].
\bibitem{KL}
D.S. Kaparulin, S.L. Lyakhovich,\\ {\it On the stability of a nonlinear oscillator with higher derivatives},\\ Russ. Phys. J. {\bf 57} (2015) 1561-1565.
\bibitem{Masterov_N2PU}
I. Masterov,\\ {\it N=2 supersymmetric Pais-Uhlenbeck oscillator},\\ Mod. Phys. Lett. A {\bf 30} (2015) 1550107, arXiv:1503.03699 [hep-th].
\bibitem{Smilga_1}
D. Robert, A.V. Smilga,\\ {\it Supersymmetry vs ghosts},\\ J. Math. Phys. {\bf 49} (2008) 042104, [math-ph/0611023].
\bibitem{Masterov_1}
I. Masterov,\\ {\it New realizations of N=2 l-conformal Newton-Hooke superalgebra},\\ Mod. Phys. Lett. A {\bf 30} (2015) 1550073, arXiv:1412.1751 [hep-th].
\bibitem{Gomis}
J. Gomis, K. Kamimura,\\ {\it Schrodinger equations for higher order non-relativistic particles and N-Galilean conformal symmetry},\\ Phys. Rev. D {\bf 85} (2012) 045023, arXiv:1109.3773 [hep-th].
\bibitem{Galajinsky_1}
A. Galajinsky, I. Masterov,\\ {\it Dynamical realizations of l-conformal Newton-Hooke group},\\ Phys. Lett. B {\bf 723} (2013) 190-195, arXiv:1303.3419 [hep-th].
\bibitem{PU}
K. Andrzejewski, A. Galajinsky, J. Gonera, I. Masterov,\\ {\it Conformal Newton-Hooke symmetry of Pais-Uhlenbeck oscillator},\\ Nucl. Phys. B {\bf 885} (2014) 150-162, arXiv:1402.1297 [hep-th].
\bibitem{Andrzejewski}
K. Andrzejewski,\\ {\it Conformal Newton-Hooke algebras, Niederer's transformation and Pais-Uhlenbeck oscillator},\\ Phys. Lett. B {\bf 738} (2014) 405-411, arXiv:1409.3926 [hep-th].
\bibitem{Andrzejewski_1}
K. Andrzejewski,\\ {\it Hamiltonian formalisms and symmetries of the Pais-Uhlenbeck oscillator},\\ Nucl. Phys. B {\bf 889} (2014) 333-350, arXiv:1410.0479 [hep-th].
\bibitem{Masterov_2}
I. Masterov,\\ {\it Dynamical realizations of N=1 l-conformal Galilei superalgebra},\\ J. Math. Phys. {\bf 55} (2014) 102901, arXiv:1407.1438 [hep-th].
\bibitem{AB}
A. Galajinsky, I. Masterov,\\ {\it On dynamical realizations of l-conformal Galilei and Newton-Hooke algebras},\\ Nucl. Phys. B {\bf 896} (2015) 244-254, arXiv:1503.08633 [hep-th].
\bibitem{Andr}
K. Andrzejewski,\\ {\it Generalized Niederer's transformation for quantum Pais-Uhlenbeck oscillator},\\ Nucl. Phys. B {\bf 901} (2015) 216-228, arXiv:1506.04909 [hep-th].

\bibitem{Negro_1}
J. Negro, M.A. del Olmo, A. Rodriguez-Marco,\\ {\it Nonrelativistic conformal groups},\\ J. Math. Phys. {\bf 38} (1997), 3786-3809.
\bibitem{Negro_2}
J. Negro, M.A. del Olmo, A. Rodriguez-Marco,\\ {\it Nonrelativistic conformal groups. II. Further developments and physical applications},\\ J. Math. Phys. {\bf 38} (1997), 3810-3831.
\bibitem{Galajinsky_3}
A. Galajinsky, I. Masterov,\\ {\it Remark on l-conformal extension of the Newton-Hooke algebra},\\ Phys.Lett. B {\bf 702} (2011) 265-267, arXiv:1104.5115 [hep-th].

\bibitem{Liu}
F.-L. Liu, Y. Tian,\\ {\it Acceleration extended Newton-Hooke symmetry and its dynamical realization},\\ Phys. Lett. A {\bf 372} (2008) 6041-6046, arXiv:0806.1310 [hep-th].


\bibitem{Lukier_1}
J. Lukierski, P.C. Stichel, W.J. Zakrzewski,\\ {\it Galilean invariant (2+1)-dimensional models with a Chern-Simons-like term and D=2 noncommutative geometry}, Annals Phys. {\bf 260} (1997) 224-249, [hep-th/9612017].
\bibitem{Horvathy}
P.A. Horvathy, M.S. Plyushchay,\\ {\it Nonrelativistic anyons in external electromagnetic field},\\ Nucl. Phys. B {\bf 714} (2005) 269-291, [hep-th/0502040].
\bibitem{Lukier_2}
J. Lukierski, P.C. Stichel, W.J. Zakrzewski,\\ {\it Acceleration-enlarged symmetries in nonrelativistic space-time with a cosmological constant},\\ Eur. Phys. J. C {\bf 55} (2008) 119-124, arXiv:0710.3093 [hep-th].


\bibitem{Horvathy_1}
P.A. Horvathy,\\ {\it Mathisson' spinning electron: Noncommutative mechanics and exotic Galilean symmetry, 66 years ago},\\ Acta Phys. Polon. B {\bf 34} (2003) 2611-2622, [hep-th/0303099].
\bibitem{Lukier_4}
J. Lukierski, P.C. Stichel, W.J. Zakrzewski,\\ {\it Exotic Galilean conformal symmetry and its dynamical realizations},\\ Phys. Lett A {\bf 357} (2006) 1, [hep-th/0511259].
\bibitem{Lukier_3}
J. Lukierski, P.C. Stichel, W.J. Zakrzewski,\\ {\it Acceleration-extended Galilean symmetries with central charges and their dynamical realizations},\\ Phys. Lett. B {\bf 650} (2007) 203-207, [hep-th/0702179].
\bibitem{Gomis_1}
P.D. Alvarez, J. Gomis, K. Kamimura, M.S. Plyushchay,\\ {\it (2+1)D exotic Newton-Hooke symmetry, duality and projective phase},\\ Ann. Phys. {\bf 322} (2007) 1556-1586, [hep-th/0702014].

\bibitem{Niederer}
U. Niederer,\\ {\it The maximal kinematical invariance group of the harmonic oscillator},\\ Helv. Phys. Acta {\bf 46} (1973), 191-200.


\bibitem{MP}
J. Lopez-Sarrion, C.M. Reyes,\\{\it Myers-Pospelov model as an ensemble of Pais-Uhlenbeck oscillators: unitarity and Lorentz invariance violation}, Eur. Phys. J. C {\bf 73} (2013) 2391, arXiv:1304.4966[hep-th].
\bibitem{Lozano}
C.S. Lozano, O. Piguet, F.A. Schaposnik, L. Sourrouille,\\{\it Nonrelativistic supersymmetry in noncommutative space},\\Phys. Lett. B {\bf 630} (2005) 108-114, [hep-th/0508009].

\bibitem{Dima_2}
D.S. Kaparulin, I. Yu. Karataeva, S.L. Lyakhovich,\\ {\it Higher derivative extensions of 3d Chern-Simons models: conservation laws and stability},\\ Eur. Phys. J. C {\bf 75} (2015) 552, arXiv:1510.02007 [hep-th].
\bibitem{N=2mb}
A. Galajinsky, I. Masterov,\\ {\it Remark on quantum mechanics with N=2 Schr\"{o}dinger supersymmetry},\\ Phys. Lett. B {\bf 675} (2009) 116-122, arXiv:0902.2910 [hep-th].
\bibitem{Galajinsky_6}
A. Galajinsky,\\ {\it N=2 superconformal Newton-Hooke algebra and many-body mechanics},\\ Phys. Lett. B {\bf 680} (2009) 510-515, arXiv:0906.5509 [hep-th].
\bibitem{harmN=2}
A. Galajinsky, O. Lechtenfeld,\\ {\it Harmonic N=2 mechanics},\\ Phys. Rev. D {\bf 80} (2009) 065012, arXiv:0907.2242 [hep-th].
\bibitem{Galajinsky_4}
A. Galajinsky,\\{\it Conformal mechanics in Newton-Hooke spacetime},\\ Nucl. Phys. B {\bf 832} (2010) 586-604, arXiv:1002.2290 [hep-th].
\bibitem{weak}
A. Smilga,\\{\it Weak supersymmetry},\\ Phys. Lett. B {\bf 585} (2004) 173-179, [hep-th/0311023].
\bibitem{Masterov_3}
I. Masterov,\\{\it Higher-derivative mechanics with N=2 l-conformal Galilei supersymmetry},\\ J. Math. Phys. {\bf 56} (2015) 022902, arXiv:1410.5335 [hep-th].

\bibitem{idty_1}
V.P. Frolov, P. Krtou\v{s}, D. Kubiz\v{n}\'{a}k,\\{\it Separability of Hamilton-Jacobi and Klein-Gordon equations in general Kerr-NUT-AdS spacetimes},\\ JHEP {\bf 0702}, (2007) 005, [hep-th/0611245].
\end{thebibliography}
\end{document}